# AXIAL QUADRUPOLE PHASE OF A UNIAXIAL SPIN-1 MAGNET

## I.P. Shapovalov[1] and P.A. Sayko[2]

I.I. Mechnikov Odesa National University, 2, Dvoryans'ka Str., Odesa, 65082, Ukraine

[1] E-mail: shapovalov@onu.edu.ua
[2] E-mail: fesswolf10@gmail.com

**Abstract.** Uniaxial phase of the axial quadrupole spin-1 magnet in an external magnetic field is investigated. A case where the system includes single-ion anisotropy and anisotropic biquadratic exchange interaction the most general form, is considered. It is proved, that the molecular field approximation of the relative magnetization does not depend on temperature and increases linearly with the external magnetic field. Two branches of the spin excitation spectrum are derived. Condition of mitigation of the spectrum defines the boundary between the axial quadrupole and angular phases. The critical temperature of the corresponding phase transition depends considerably on the anisotropy constants of the biquadratic exchange interaction.

# ОСЕВАЯ КВАДРУПОЛЬНАЯ ФАЗА ОДНООСНОГО СПИН-1 МАГНЕТИКА

**И.П. Шаповалов[1], П.А. Сайко[2]**

*Одесский национальный университет им. И.И. Мечникова
ул. Дворянская, 2, г. Одесса, 65082, Украина*

[1] E-mail: shapovalov@onu.edu.ua
[2] E-mail: fesswolf10@gmail.com

Исследована осевая квадрупольная фаза одноосного спин-1 магнетика во внешнем магнитном поле. Рассмотрен случай, когда в системе присутствуют одноионная анизотропия и анизотропное биквадратичное обменное взаимодействие наиболее общего вида. Доказано, что в приближении молекулярного поля относительная намагниченность не зависит от температуры и линейно возрастает при увеличении внешнего магнитного поля. Определены две ветви спектра спиновых возбуждений. Условие смягчения спектра задаёт границу между осевой квадрупольной и угловой фазами. Критическая температура соответствующего фазового перехода существенно зависит от констант анизотропии биквадратичного обменного взаимодействия.

Досліджено осьову квадрупольну фазу одновісного спін-1 магнетика у зовнішньому магнітному полі. Розглянуто випадок, коли в системі присутні одноіонна анізотропія і анізотропна біквадратична обмінна взаємодія найбільш загального вигляду. Доведено, що у наближенні молекулярного поля відносна намагніченість не залежить від температури та лінійно зростає зі збільшенням зовнішнього магнітного поля. Визначено дві вітки спектра спінових збуджень. Умова пом'якшення спектра задає межу між осьовою квадрупольною та кутовою фазами. Критична температура відповідного фазового переходу суттєво залежить від констант анізотропії біквадратичної обмінної взаємодії.

Ключевые слова: одноионная анизотропия, биквадратичное обменное взаимодействие, квадрупольная фаза, граница устойчивости.

## 1. Введение

Исследование сильных магнетиков со спином ½ традиционно проводится на основе модели Гейзенберга. Эта модель включает в себя оператор энергии обменного взаимодействия (ОВ), имеющий билинейную по спиновым операторам форму.

В магнитных системах со спином, значение которого превосходит ½, могут возникать дополнительные взаимодействия, выражающиеся через компоненты тензора квадрупольного магнитного момента. При единичном значении спина, которое рассматривается в настоящей работе, тензорные взаимодействия можно разделить на два типа: одноионная анизотропия (ОА) и биквадратичное обменное взаимодействие (БОВ). ОА имеет форму линейную по квадрупольным операторам и, соответственно, квадратичную по спиновым операторам. БОВ, являясь двухчастичным, представляет собой билинейную по квадрупольным и биквадратичную по спиновым операторам форму.

В тех случаях, когда тензорные взаимодействия малы, модель Гейзенберга оказывается приемлемой для исследования магнетиков со спином большим ½. Однако существует множество магнитных соединений, для которых константы тензорных взаимодействий по порядку величины сравнимы с константой ОВ [1]. Для адекватного описания свойств таких веществ модель Гейзенберга необходимо расширить путём включения в неё тензорных взаимодействий.

Наличие большой ОА и сильного БОВ может быть причиной возникновения в системе так называемых квадрупольных фаз. Характерная особенность квадрупольных фаз состоит в том, что в нулевом магнитном поле намагниченность системы равна нулю, а квадрупольная составляющая параметра порядка отлична от нуля [1].

В системах с единичным спином существует два различных типа квадрупольного упорядочения. При упорядочении первого типа в отсутствие внешнего магнитного поля при нулевой температуре выполняется условие: $\langle S^\alpha \rangle = 0$, $\langle (S^\alpha)^2 \rangle = 1$, где $S^\alpha$ – проекция магнитного момента иона на ось упорядочения α [1]. Фазы такого типа мы будем обозначать символом КУ$_{1\alpha}$. В этих фазах величина $S^\alpha$ с равной вероятностью может принимать значения 1 и −1, т.е. фазы типа КУ$_{1\alpha}$ являются осевыми квадрупольными фазами. Второй тип квадрупольного упорядочения соответствует условию $\langle S^\alpha \rangle = 0$; $\langle (S^\alpha)^2 \rangle = 0$ [1]. Фазы с таким упорядочением мы будем обозначать символом КУ$_{2\alpha}$. В этих фазах магнитные моменты ионов свободно вращаются в плоскости, перпендикулярной оси α. Иными словами, фазы типа КУ$_{2\alpha}$ являются плоскостными квадрупольными фазами.

В научной литературе накоплен значительный материал по исследованию свойств магнитных систем с ОА и изотропным БОВ [2-15]. При этом особенности поведения магнетиков с анизотропным БОВ изучены недостаточно. Следует отметить, что те немногие работы, в которых учитывалась анизотропия БОВ, показывают, что изменение констант анизотропии может приводить к существенному изменению свойств магнитной системы [16-20].

В [19] изучены возможные квадрупольные фазы в одноосном спин-1 магнетике при наличии ОА и анизотропного БОВ наиболее общего вида в отсутствие внешнего магнитного поля. В общем случае возможны три квадрупольные фазы: КУ$_{2Z}$, КУ$_{2X}$ и КУ$_{2<}$. В фазе КУ$_{2Z}$ в основном состоянии проекция магнитного момента каждого иона на ось симметрии (ось OZ) равна нулю, т.е. плоскостью упорядочения является плоскость XOY. Плоскость упорядочения, соответствующая фазе КУ$_{2X}$, – это плоскость YOZ. В фазе КУ$_{2<}$ плоскость упорядочения включает в себя ось OY и образует с осью OZ некоторый угол, который непрерывно изменяется при изменении параметров гамильтониана. Для квадрупольной фазы КУ$_{2X}$ определены две ветви спектра спиновых возбуждений в при конечных температурах. Граница устойчивости спектра совпадает с линией фазовых переходов второго рода из фазы КУ$_{2X}$ в фазу КУ$_{2<}$ (если последняя реализуется в системе).

Важно отметить, что в фазе КУ$_{2X}$ совместно с условием $\langle S^X \rangle = 0$; $\langle (S^X)^2 \rangle = 0$ выполняется условие $\langle S^Z \rangle = 0, \langle (S^Z)^2 \rangle = 1$. Таким образом, фаза КУ$_{2X}$ является одновременно фазой второго типа относительно оси OX и фазой первого типа относительно оси OZ, т.е. фазой КУ$_{1Z}$. Однако такая ситуация реализуется только в нулевом поле. Включение внешнего поля, направленного вдоль оси симметрии OZ, приводит к разрушению свободного вращения магнитных моментов ионов в плоскости XOY. Соответственно, обозначение КУ$_{2X}$ оказывается неудачным, и мы будем использовать обозначение КУ$_{1Z}$. Наличие поля приводит также к тому, что при измерении величины $S^Z$ вероятность получить значение 1 оказывается больше ½. Эта вероятность увеличивается при увеличении поля.

Целью настоящей работы является обобщение результатов, полученных в [19], на случай наличия внешнего магнитного поля.

## 2. Температурная и полевая зависимость параметра порядка

Гамильтониан одноосного спин-1 магнетика при наличии ОА и БОВ в наиболее общем случае имеет вид [16]

$$H = -h\sum_i S_i^Z - \sum_{i,j} J_{ij}\left[S_i^Z S_j^Z + \xi\left(S_i^X S_j^X + S_i^Y S_j^Y\right)\right] + \frac{1}{3}D\sum_i Q_i^0 - \\ -\sum_{i,j} K_{ij}\left[\frac{1}{3}Q_i^0 Q_j^0 + \eta\left(Q_i^1 Q_j^1 + Q_i^{-1} Q_j^{-1}\right) + \varsigma\left(Q_i^2 Q_j^2 + Q_i^{-2} Q_j^{-2}\right)\right], \quad (1)$$

где $h$ – напряжённость внешнего магнитного поля, направленного вдоль оси OZ; $J_{ij}$ – константы ОВ; $\xi$ – константа анизотропии ОВ; $D$ – константа ОА; $K_{ij}$ – константы БОВ; $\eta, \varsigma$ – константы анизотропии БОВ; $S^l$ ($l = x, y, z$) – спиновые операторы; $Q^p$ ($p = 0, \pm 1, \pm 2$) – квадрупольные операторы; индексы $i$, $j$ нумеруют узлы кристаллической решётки.

Связь операторов $Q^p$ со спиновыми операторами задаётся следующим образом:

$$Q^0 = 3(S^z)^2 - 2I, \quad Q^1 = -(S^z S^y + S^y S^z), \quad Q^{-1} = S^z S^x + S^x S^z,$$
$$Q^2 = (S^x)^2 - (S^y)^2, \quad Q^{-2} = -(S^x S^y + S^y S^x),$$

где $I$ – единичная матрица.

Восемь операторов $S^l$, $Q^p$ образуют базис алгебры Ли группы SU(3). При этом операторы $S^Z$ и $Q^0$ являются диагональными. Матричный вид операторов $S^l$, $Q^p$ приведен в Приложении.

В том случае, когда выполняются неравенства $J_{ij} > 0$ и $K_{ij} > 0$, в системе могут реализовываться только одноподрешёточные фазы. При этом средние значения $\langle S^l \rangle$ и $\langle Q^p \rangle$ полностью определят спиновую структуру системы и, следовательно, могут быть использованы в качестве компонент параметра порядка (ПП).

В отсутствие внешнего магнитного поля фаза осевого квадрупольного упорядочения КУ$_{1Z}$ имеет две отличные от нуля компоненты ПП: $\langle Q^0 \rangle$ и $\langle Q^2 \rangle$. Включение поля приводит к появлению в системе продольной составляющей намагниченности: $\langle S^z \rangle \neq 0$. Соответственно, в приближении молекулярного поля гамильтониан системы имеет вид

$$H_0 = -\left(h + 2J_0 \langle S^z \rangle\right)\sum_i S_i^z + \frac{1}{3}\left(D - 2K_0 \langle Q^0 \rangle\right)\sum_i Q_i^0 - 2\varsigma K_0 \langle Q^2 \rangle \sum_i Q_i^2 \; , \qquad (2)$$

где величины $J_0$ и $K_0$ – определяются равенствами

$$J_0 = \sum_i J_{ij}, \; K_0 = \sum_i K_{ij} \; .$$

Произвольный одночастичный гамильтониан спин-1 магнетика может быть диагонализован с помощью унитарного преобразования базисных операторов $S^l$ и $Q^p$:

$$S^l = V \widetilde{S}^l V^{-1}; \; Q^p = V \widetilde{Q}^p V^{-1}, \qquad (3)$$

где $\widetilde{S}^l$ и $\widetilde{Q}^p$ – новые базисные операторы, $V$ – подходящее унитарное преобразование. Общий вид преобразования $V$ получен в [21]:

$$V = \exp\left(-i\psi\, Q^{-2}\right) \times \exp\left(i\chi\, Q^1\right) \times \exp\left(i\varphi\, S^y\right), \qquad (4)$$

где $\varphi, \chi, \psi$ – параметры преобразования.

Для гамильтониана (2) параметры преобразования определяются следующим образом [16]:

$$\varphi = -\pi/2, \; \psi = \pi/4, \; \sin 2\chi = \frac{-h}{(K_0\varsigma - J_0)(\sigma + \lambda)}, \qquad (5)$$

где введены обозначения: $\sigma \equiv \langle \widetilde{S}^z \rangle$, $\lambda \equiv \langle \widetilde{Q}^0 \rangle$.

Величины $\sigma$ и $\lambda$ зависят от параметров гамильтониана и температуры.

Усредняя соотношения (3) с учётом явного вида преобразования $V$, можно найти отличные от нуля компоненты ПП как функции параметров гамильтониана и величин $\sigma$ и $\lambda$:

$$<S^z> = \frac{h}{2(K_0\varsigma - J_0)}, \; \langle Q^0 \rangle = (3\sigma - \lambda)/2, \; \langle Q^2 \rangle = -\frac{\sqrt{(K_0\varsigma - J_0)^2(\sigma + \lambda)^2 - h^2}}{2(K_0\varsigma - J_0)}. \qquad (6)$$

В нулевом поле формулы (6) совпадают с соответствующими формулами работы [19].

Из первого равенства (6) следует интересный результат: в фазе КУ$_{1Z}$ в приближении молекулярного поля относительная намагниченность не зависит от температуры и линейно растёт с ростом внешнего магнитного поля.

В новом базисе гамильтониан $H_0$ имеет диагональный вид:

$$H_0 = \widetilde{\alpha}_1 \sum_i \widetilde{S}_i^Z + \widetilde{\alpha}_2 \sum_i \widetilde{Q}_i^0 \,, \qquad (7)$$

где величины $\widetilde{\alpha}_1$ и $\widetilde{\alpha}_2$ определяются следующими выражениями:

$$\widetilde{\alpha}_1 = \frac{1}{2}D - 2\sigma K_0 - \frac{1}{2}K_0(\varsigma-1)(\sigma+\lambda)\,, \quad \widetilde{\alpha}_2 = -\frac{1}{6}D - \frac{2}{3}\lambda K_0 - \frac{1}{2}K_0(\varsigma-1)(\sigma+\lambda)\,. \qquad (8)$$

Энергетические уровни отдельного иона, соответствующие различным значениям величины $\widetilde{S}^z$ ($\widetilde{S}^z = 1, 0, -1$):

$$E_1 = \widetilde{\alpha}_1 + \widetilde{\alpha}_2\,, \quad E_0 = -2\widetilde{\alpha}_2\,, \quad E_{-1} = -\widetilde{\alpha}_1 + \widetilde{\alpha}_2\,. \qquad (9)$$

Величины $\sigma$ и $\lambda$ можно определить с помощью распределения Гиббса:

$$\sigma = \frac{\sum_n \widetilde{S}_n^Z \exp(-E_n/kT)}{\sum_n \exp(-E_n/kT)}\,, \qquad \lambda = \frac{\sum_n \widetilde{Q}_n^Z \exp(-E_n/kT)}{\sum_n \exp(-E_n/kT)}\,, \qquad (10)$$

где $k$ – постоянная Больцмана, $T$ – абсолютная температура.

В дальнейшем мы будем использовать такое решение системы (10), для которого при стремлении к нулю температуры $T$ выполняется условие $\sigma = 0$, $\lambda = -2$, т.е. нижним энергетическим уровнем иона является уровень с $\widetilde{S}^z = 0$ [19].

Из формул (8-10) следует, что величины $\widetilde{\alpha}_1$ и $\widetilde{\alpha}_2$, энергетические уровни отдельного иона $E_n$ и величины $\sigma$ и $\lambda$ не зависят от внешнего магнитного поля.

Отсутствие полевой зависимости величин $\sigma$ и $\lambda$ приводит к тому, что в фазе КУ$_{1Z}$ квадрупольная составляющая ПП $\langle Q^0 \rangle$, задаваемая второй формулой (6), не зависит от поля $h$. Зависимость величины $\langle Q^0 \rangle$ от безразмерной температуры $\widetilde{\theta}$ $(\widetilde{\theta} = \theta/J_0)$ приведена на рисунке 1. Следует отметить, что с увеличением поля происходит фазовый переход (ФП) второго рода из фазы КУ$_{1Z}$ в ферромагнитную фазу (ФМФ) [22]. Критическая температура $\widetilde{\theta}_C$ этого ФП, т.е. температура окончания линии $\langle Q^0 \rangle = f(\widetilde{\theta})$, зависит от величины $h$. На рисунке величина $\widetilde{\theta}_C$ соответствует значению безразмерного магнитного поля $\widetilde{h} = 4{,}5$ $\left(\widetilde{h} = h/J_0\right)$.

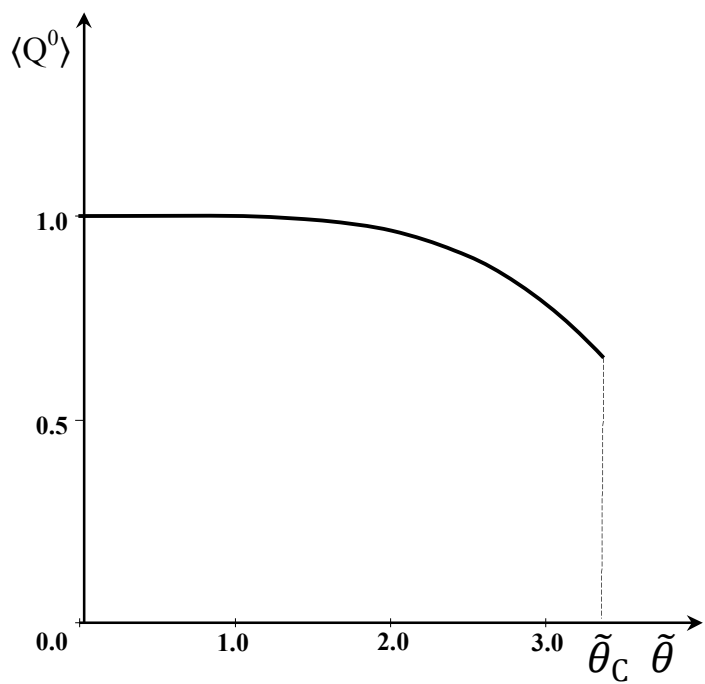

Рис.1. Температурная зависимость квадрупольной компоненты ПП $\langle Q^0 \rangle$. Линия построена при $J_0$=1.0; ξ=1.0; D=1.2; K$_0$=1.25; η=2.0; ζ=3.0.

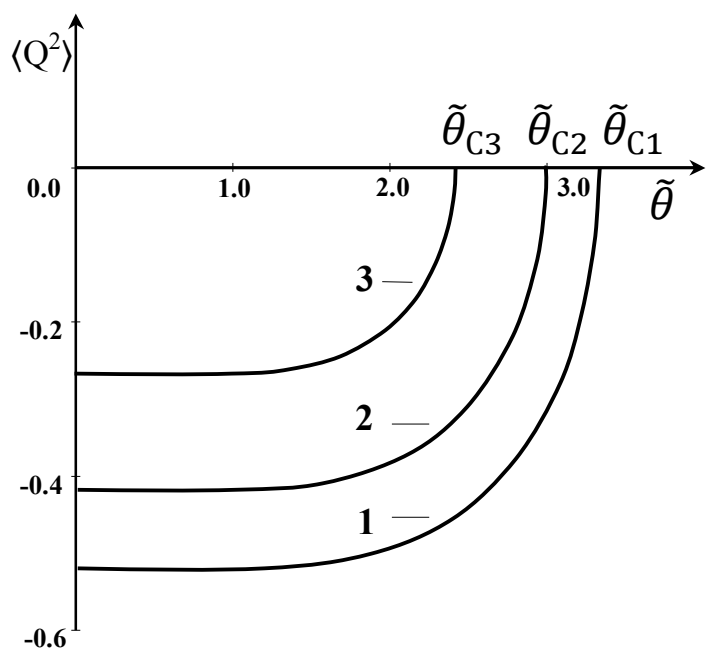

Рис.2. Температурная зависимость квадрупольной компоненты ПП $\langle Q^2 \rangle$: 1 – $\widetilde{h} = 4{,}7$; 2 – $\widetilde{h} = 5{,}0$; 3 – $\widetilde{h} = 5{,}3$. Линии построены при $J_0$=1.0; ξ=1.0; D=1.2; K$_0$=1.25; η=2.0; ζ=3.0.

На рисунке 2 представлена зависимость величины $\langle Q^2 \rangle$ от безразмерной температуры при различных значениях внешнего магнитного поля. Поскольку в ФМФ выполняется условие $\langle Q^2 \rangle = 0$, точки $\tilde{\theta}_{Cl}$ ($l = 1, 2,...$) для каждой кривой на рисунке 2 являются точками ФП в ФМФ. Таким образом, семейство кривых на рисунке 2 позволяет определить множество точек фазовой границы КУ$_{1Z}$ $\leftrightarrow$ ФМФ в координатах $\tilde{h} - \tilde{\theta}$. Эта граница приведена на рисунке 3. Она полностью совпадает

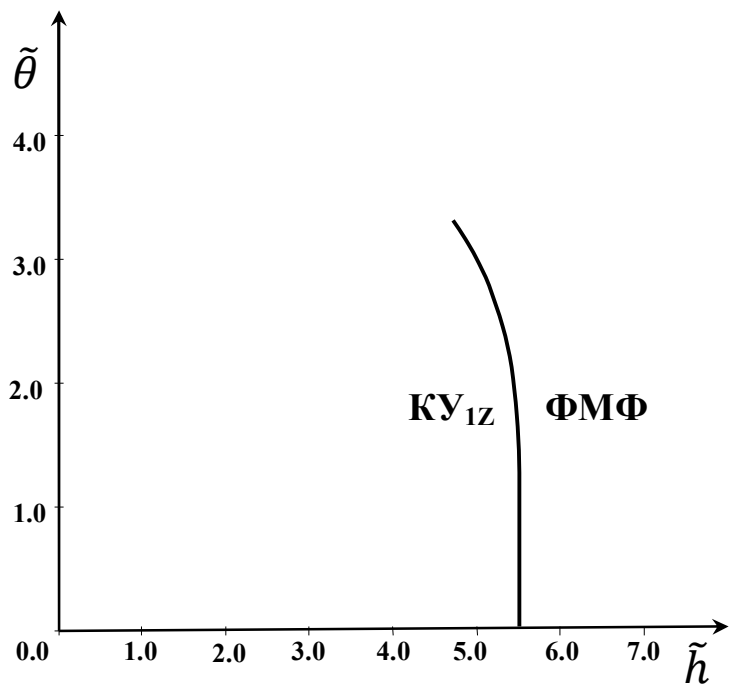

Рис.3. Граница между фазой КУ$_{1Z}$ и ФМФ. Линия построена при $J_0$=1.0; $\xi$=1.0; $D$=1.2; $K_0$=1.25; $\eta$=2.0; $\zeta$=3.0.

с линией потери устойчивости спектра спиновых возбуждений в ФМФ, построенной по алгоритму работы [22].

## 3. Спектр спиновых возбуждений

Поскольку нижним энергетическим уровнем отдельного иона является уровень с $\tilde{S}^z = 0$, состояния с $\tilde{S}^z = -1$ и $\tilde{S}^z = 1$ являются возбуждёнными. Две ветви спектра спиновых возбуждений при условии $h = 0$ получены в [19]:

$$\omega_1(\mathbf{k}) = \{[2(\lambda - \sigma)(\eta K_{\mathbf{k}} - K_0) - (\lambda + \sigma)K_0(\varsigma - 1) - D] \times \\ \times [2(\lambda - \sigma)(\xi J_{\mathbf{k}} - K_0) - (\lambda + \sigma)K_0(\varsigma - 1) - D]\}^{1/2}, \tag{11}$$

$$\omega_2(\mathbf{k}) = 2|\sigma + \lambda|\sqrt{\varsigma(K_0 - K_\mathbf{k})(\varsigma K_0 - J_\mathbf{k})}. \tag{12}$$

Обобщение расчётов на случай наличия внешнего магнитного поля даёт следующее выражение для первой ветви спектра:

$$\omega_1(\mathbf{k}) = \left\{ L(\mathbf{k}) + \sqrt{L^2(\mathbf{k}) - 4M(\mathbf{k})} \right\}^{1/2}, \tag{13}$$

где

$$L(\mathbf{k}) = A_\mathbf{k}^2 - B_\mathbf{k}^2 + G_\mathbf{k}^2 - H_\mathbf{k}^2 + 2C_\mathbf{k}E_\mathbf{k} - 2D_\mathbf{k}F_\mathbf{k},$$

$$M(\mathbf{k}) = \det \begin{Vmatrix} A_\mathbf{k} & B_\mathbf{k} & C_\mathbf{k} & D_\mathbf{k} \\ -B_\mathbf{k} & -A_\mathbf{k} & -D_\mathbf{k} & -C_\mathbf{k} \\ E_\mathbf{k} & F_\mathbf{k} & G_\mathbf{k} & H_\mathbf{k} \\ -F_\mathbf{k} & -E_\mathbf{k} & -H_\mathbf{k} & -G_\mathbf{k} \end{Vmatrix},$$

$$A_\mathbf{k} = h\sin 2\chi - D + (3\sigma - \lambda)K_0 - (\sigma + \lambda)(J_0 \sin^2 2\chi + \varsigma K_0 \cos^2 2\chi) - $$
$$- (\sigma - \lambda)(\xi J_\mathbf{k} + \eta K_\mathbf{k}),$$
$$B_\mathbf{k} = (\sigma - \lambda)\cos 2\chi (\xi J_\mathbf{k} - \eta K_\mathbf{k}),$$
$$C_\mathbf{k} = h\cos 2\chi - (\sigma + \lambda)\sin 2\chi \cos 2\chi (J_0 - \varsigma K_0),$$
$$D_\mathbf{k} = (\sigma - \lambda)\sin 2\chi (\eta K_\mathbf{k} - \xi J_\mathbf{k}),$$
$$E_\mathbf{k} = C_\mathbf{k},$$
$$F_\mathbf{k} = 2\sigma \sin 2\chi (\eta K_\mathbf{k} - \xi J_\mathbf{k}),$$
$$G_\mathbf{k} = -h\sin 2\chi - D + (3\sigma - \lambda)K_0 + (\sigma + \lambda)(J_0 \sin^2 2\chi + \varsigma K_0 \cos^2 2\chi) - $$
$$- 2\sigma(\xi J_\mathbf{k} + \eta K_\mathbf{k}),$$
$$H_\mathbf{k} = 2\sigma \cos 2\chi (\eta K_\mathbf{k} - \xi J_\mathbf{k}).$$

Условие устойчивости ветви спектра $\omega_1(\mathbf{k})$ имеет вид $\omega_1(0) > 0$, а граница устойчивости:

$$\omega_1(0) = 0. \tag{14}$$

Равенство (14) эквивалентно следующей системе:

$$L(0) < 0, \tag{15}$$

$$M(0) = 0. \tag{16}$$

Вторая ветвь спектра имеет вид

$$\omega_2(\mathbf{k}) = 2\{[h\sin 2\chi + (\sigma + \lambda)(\varsigma K_\mathbf{k} - J_0)\sin^2 2\chi - (\sigma + \lambda)(\varsigma K_0 - J_\mathbf{k})\cos^2 2\chi] \times$$
$$\times [h\sin 2\chi + (\sigma + \lambda)(\varsigma K_\mathbf{k} - J_0)\sin^2 2\chi - \varsigma(\sigma + \lambda)(K_0 - K_\mathbf{k})\cos^2 2\chi]\}^{1/2}. \tag{17}$$

При нулевой температуре выражения (13) и (17) полностью согласуются с результатами работы [16], а при конечных температурах и нулевом поле – с результатами работы [19].

Отметим, что при условии $h=0$ мода ветви спектра $\omega_2(\mathbf{k})$ с нулевым значением волнового вектора является модой Голдстоуна. При конечных значениях поля частота мягкой моды отлична от нуля.

В [19] показано, что в нулевом поле условие (14) описывает границу между осевой квадрупольной фазой и фазой КУ$_<$. При условии $h \neq 0$ фаза КУ$_<$ в системе не реализуется, а условие (14) задаёт линию ФП второго рода между фазой КУ$_{1Z}$ и угловой фазой, в которой намагниченность образует определённый угол с осью OZ [16, 18]. Критическая температура $\theta_{кр}$ этих ФП является неявной функцией констант анизотропии БОВ $\eta$ и $\varsigma$. Зависимость безразмерной критической

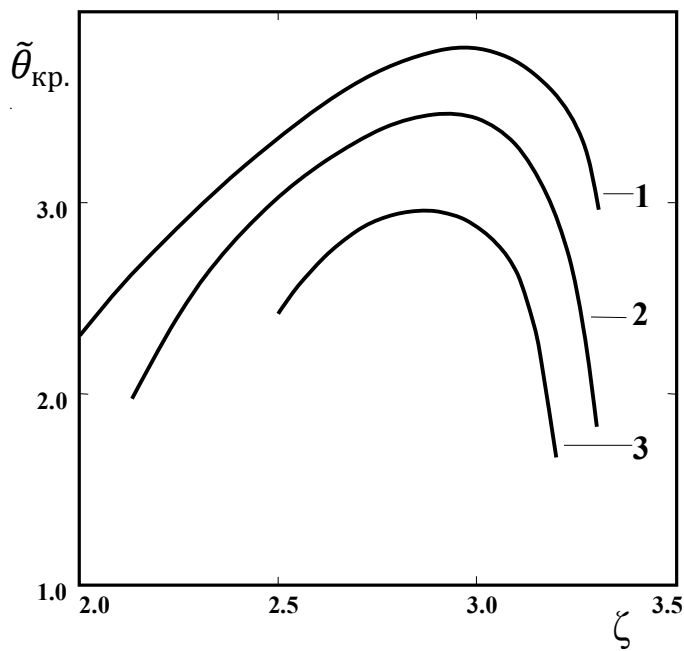

Рис.4. Зависимость безразмерной критической температуры $\widetilde{\theta}_{кр.}$ от величины $\varsigma$ при различных значениях безразмерного поля: 1 – $\widetilde{h}=2.5$; 2 – $\widetilde{h}=3.2$; 3 – $\widetilde{h}=3.8$. Линии построены при $J_0=1.0$; $\xi=1.0$; $D=1.2$; $K_0=1.25$; $\eta=2.0$.

температуры $\widetilde{\theta}_{кр}$ от величин $\eta$ и $\varsigma$ при различных значениях безразмерного поля $\widetilde{h}$ приведена на рисунках 4 и 5.

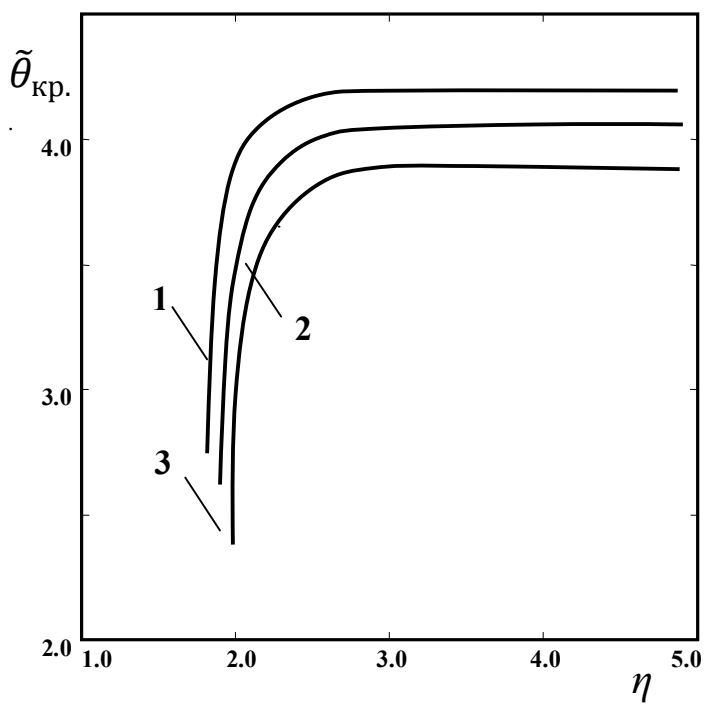

Рис.5. Зависимость безразмерной критической температуры $\tilde{\theta}_{кр.}$ от величины $\eta$ при различных значениях безразмерного поля: 1 – $\tilde{h} = 2.5$; 2 – $\tilde{h} = 3.2$; 3 – $\tilde{h} = 3.8$. Линии построены при $J_0=1.0$; $\xi=1.0$; $D=1.2$; $K_0=1.25$; $\zeta =3.0$.

## 4. Обсуждение результатов

Как показано выше, в осевой квадрупольной фазе КУ$_{1Z}$ в приближении молекулярного поля выражения для энергетических уровней отдельного иона не зависят от величины внешнего магнитного поля. Это приводит к отсутствию полевой зависимости выражения для энергии основного состояния системы и, как следствие, его совпадению с соответствующим выражением работы [19]. В плоскостной квадрупольной фазе КУ$_{2Z}$ выражение для энергии основного состояния также не включает в себя поле $h$ [16]. Поэтому при нулевой температуре ФП по полю между осевой и плоскостной квадрупольными фазами невозможны. Таким образом, если в системе при нулевой температуре возникла фаза КУ$_{1Z}$, а угловая фаза в системе не реализуется, то при уменьшении поля фаза КУ$_{1Z}$ будет сохраняться вплоть до значения поля $h = 0$.

Если в системе возможна реализация угловой фазы, то температура ФП из фазы КУ$_{1Z}$ в угловую фазу сильно зависит от констант анизотропии БОВ $\eta$ и $\varsigma$. Это обстоятельство свидетельствует в пользу таких моделей одноосного спин-1 магнетика, в которых БОВ учитывается в анизотропном виде.

С увеличением магнитного поля $h$ неизбежно происходит ФП второго рода из фазы КУ$_{1Z}$ в ФМФ. Граница обеих фаз в координатах поле – температура представляет собой границу устойчивости спектра спиновых возбуждений в фазе ФМФ [22]. Равенство нулю компоненты параметра порядка $\langle Q^2 \rangle$ в фазе ФМФ и её отличие от нуля в фазе КУ$_{1Z}$ даёт альтернативный способ определения фазовой границы КУ$_{1Z} \leftrightarrow$ ФМФ.

## Приложение.
### Матричный вид операторов $S^l$ и $Q^p$

$$S^Z = \begin{vmatrix} 1 & 0 & 0 \\ 0 & 0 & 0 \\ 0 & 0 & -1 \end{vmatrix} \quad S^X = \frac{1}{\sqrt{2}} \begin{vmatrix} 0 & 1 & 0 \\ 1 & 0 & 1 \\ 0 & 1 & 0 \end{vmatrix} \quad S^Y = \frac{i}{\sqrt{2}} \begin{vmatrix} 0 & -1 & 0 \\ 1 & 0 & -1 \\ 0 & 1 & 0 \end{vmatrix}$$

$$Q^0 = \begin{vmatrix} 1 & 0 & 0 \\ 0 & -2 & 0 \\ 0 & 0 & 1 \end{vmatrix} \quad Q^1 = \frac{i}{\sqrt{2}} \begin{vmatrix} 0 & 1 & 0 \\ -1 & 0 & -1 \\ 0 & 1 & 0 \end{vmatrix} \quad Q^{-1} = \frac{1}{\sqrt{2}} \begin{vmatrix} 0 & 1 & 0 \\ 1 & 0 & -1 \\ 0 & -1 & 0 \end{vmatrix}$$

$$Q^2 = \begin{vmatrix} 0 & 0 & 1 \\ 0 & 0 & 0 \\ 1 & 0 & 0 \end{vmatrix} \quad O_2^{-2} = i \begin{vmatrix} 0 & 0 & 0 \\ 0 & 0 & 0 \\ -1 & 0 & 0 \end{vmatrix}$$

1. В.Л. Нагаев, *Магнетики со сложными обменными взаимодействиями*. Наука, Москва, 1988.
2. Ф. П. Онуфриева, *ЖЭТФ* **89**, 2270 (1985).
3. В.П. Дьяконов, Э.Е. Зубов, Ф.П. Онуфриева и др., *ЖЭТФ* **93**, 1775 (1987).
4. В.В. Вальков, Б.В. Федосеев, *ФТТ* **32**, 3522 (1990).
5. В.М. Локтев, В.С. Островский, *ФНТ* **20**, 983 (1994) [*Low Temp. Phys.* **20**, 775 (1994)].
6. В.М. Калита, В.М. Локтев, *ФНТ* **28**, 1244 (2002) [*Low Temp. Phys.* **28**, 883 (2002)].
7. O.R. Baran, R.R. Levitskii, *Phys. Rev.* B **65**, 172407 (2002).
8. S.A. Zvyagin, J. Wosnitza, C.D. Batista et al., *Phys. Rev. Lett.* **98**, 047205 (2007).
9. V.S. Zapf, V.F. Correa, C.D. Batista, T.P. Murphy, E.D. Palm et al., *J. Appl. Phys.* **101**, 09E106 (2007).
10. В.И. Бутрим, Б.А. Иванов, А.С.Кузнецов, Р.С. Химин, *ФНТ* **34**, 1266 (2008) [*Low Temp. Phys.* **34**, 997 (2008)].
11. V. Massidda, *J. Magn. Magn. Mater.* **320**, 851 (2008).
12. M.T. Borowiec, V.P. Dyakonov, E.N. Khatsko, T. Zayarnyuk, E.E. Zubov at al., *ФНТ* **37**, 854 (2011) [*Low Temp. Phys.* **37**, 678 (2011)].
13. W. Selke, *Phys. Rev. E* **87**, 014101 (2013).
14. Ph.N. Klevets, O.A.Kosmachev, Yu.A.Fridman, *J. Magn. Magn. Mater.* **330**, 91 (2013).
15. E. G. Galkina, V. I. Butrim, Yu. A. Fridman et al., *Phys. Rev. B* **88**, 144420 (2013).
16. F.P. Onufrieva, I.P. Shapovalov, *J. Moscov. Phys. Soc.* **1**, 63 (1991).
17. Ю.А. Фридман, О.А. Космачев, Ф.Н. Клевец, *ФНТ* **32**, 289 (2006) [*Low Temp. Phys.* **32**, 214 (2006)].
18. Yu.A. Fridman, O.A. Kosmachev, Ph.N. Klevets, *J. Magn. Magn. Mater.* **320**, 435 (2008).
19. И.П. Шаповалов, *ФНТ* **39**, 663 (2013).
20. I.P. Shapovalov, P.A. Sayko, *J. Magn. Magn. Mater.* **348**, 132 (2013).
21. Ф.П. Онуфриева, *ЖЭТФ* **80**, 2372 (1981).
22. І. П. Шаповалов, *УФЖ* **53**, 653 (2010).